# Nanosecond mid-infrared pulse generation via modulated thermal emissivity

Yuzhe Xiao[1], Nicholas A. Charipar[2], Jad Salman[1], Alberto Piqué[2] and Mikhail A. Kats[1,3,4,*]

[1]Department of Electrical and Computer Engineering, University of Wisconsin-Madison, Madison, Wisconsin, USA 53706
[2]Naval Research Laboratory, 4555 Overlook Ave. SW, Washington, DC 20375, USA
[3]Materials Science and Engineering, University of Wisconsin-Madison, Madison, Wisconsin, USA 5370
[4]Department of Physics, University of Wisconsin - Madison, Madison, Wisconsin 53706, USA
*Email address: mkats@wisc.edu

We demonstrate the generation of nanosecond mid-infrared pulses via fast modulation of thermal emissivity, enabled by the absorption of visible pump pulses in unpatterned silicon and gallium arsenide. The free-carrier dynamics in these materials result in nanosecond-scale modulation of thermal emissivity, leading to nanosecond pulsed thermal emission. To our knowledge, the nanosecond thermal-emissivity modulation demonstrated in this work is three orders of magnitude faster than what has been previously demonstrated. We also indirectly observed sub-nanosecond thermal pulses from hot carriers in semiconductors. The experiments are well described by our multiphysics model. Our method of converting visible pulses into the mid infrared using modulated emissivity obeys different scaling laws and can have significant wavelength tunability compared to approaches based on conventional nonlinearities.

Short optical pulses have applications ranging from telecommunication to ultrafast science to materials processing. Despite the relative maturity of pulsed sources in the visible and near-infrared spectral ranges, there is a deficiency of sources operating in the mid infrared (wavelengths $\sim 2 - 20$ μm). Existing technologies have significant limitations; for example, mode locking of quantum cascade lasers is challenging and has resulted in sources with low power and limited tunability [1] [2], while down conversion of near-infrared pulses using nonlinear optics requires complex and expensive instrumentation [3] [4]. Here, we explore an approach for generating short pulses in the mid infrared based on fast optically driven modulation of thermal emission.

According to Planck's and Kirchhoff's laws, the optical power thermally emitted by an object depends both on its temperature and emissivity [5]. Modulation of thermal emission can therefore be realized via dynamic changes in either of these two parameters. Fast modulation of temperature is in principle possible for emitters with small volumes, and hence small heat capacity. For example, electrical heating of carbon-nanotube films has been used to demonstrate thermal-emission modulation of up to 10 GHz [6]. Even faster temperature change can be realized by decoupling the electronic and lattice temperatures: electrons can be driven far out of thermal equilibrium with phonons for a very short amount of time when pumped by a laser pulse [7]. Observations of hot-electron thermal emission have been reported in graphene [8] and in metals [9], though not in semiconductors.

While temperature-modulation speed is inherently limited by the emitter's heat capacity, the modulation of emissivity has no such restrictions. A tunable emissivity can be achieved using materials whose optical properties change in response to external factors, such as voltage [10] [11] [12] [13], optical field [14] [15], temperature [16][17][18], and strain [19][20]. The fastest demonstration of emissivity modulation thus far used carrier-density tuning in a quantum-well-based gated thermal emitter, resulting in modulation as fast as a few microseconds [13]. To the best of our knowledge, there has not been any experimental demonstration of modulation of emissivity at or below the nanosecond time scale. In this work, we demonstrate pulse generation in the mid infrared based on fast modulation of the thermal emissivity of semiconductors at time scales of just a few nanoseconds—three orders of magnitude faster than the previous record [13]. We also detected, for the first time, sub-nanosecond hot-carrier thermal emission from semiconductor emitters.

Here, mid-infrared pulses are generated by rapidly modulating thermal emission from heated semiconductors using a visible pulsed laser ($\lambda$ = 515 nm) [Fig. 1(a)]. Undoped semiconductors are generally poor thermal emitters at photon energies below their band gap (*e.g.*, silicon (Si) or gallium arsenide (GaAs) in the mid infrared [21]), but can become highly emissive via the presence of free carriers that can be generated via absorption of an above-gap optical pump pulse. Note that we do not consider the case where the free-carrier density is so high that the material becomes metallic; in this extreme regime, the emissivity can be low even with high carrier density. As can be described by the Drude model [22], an increase in free-carriers density increases the optical absorption, resulting in an increased emissivity as expected from Kirchhoff's law of thermal emission [23]. After the pump is turned off, the free carriers recombine (this process takes a few nanoseconds for GaAs [21] and a few microseconds for Si [21]), and the emissivity decreases. This mechanism of thermal-emission modulation via emissivity change through free-carrier dynamics can thus be expected to generate thermal pulses with the duration mainly determined by the free-carrier lifetime. Meanwhile, for much shorter durations (typically a few picoseconds), the free carriers can be heated up to very high temperatures before they equilibrate with the lattice [8] [9]. Thermal emission from these hot carriers is expected to have temporal widths of only a few picoseconds due to the rapid rise and fall in free-carrier temperature.

Therefore, we expect ultrafast above-gap pumping of an undoped semiconductor to result in pulsed thermal emission with two different time scales (nanosecond and picosecond); the former from the change in free-carrier density, and the latter from hot-carrier dynamics. The spectrum and temporal width of the resulting emitted pulses should depend on both the pump characteristics and the semiconductor properties. To validate this idea, we carried out both experimental and theoretical investigations. In the following, we first present the experimental data, and then build a model to explain the experimental results.

We tested our hypotheses using the experimental setup shown schematically in Fig. 1(a). The emitter—a flat unpatterned intrinsic Si (thickness of 500 µm and background doping concentration of $2 \times 10^{11}$ cm$^{-3}$) or GaAs (thickness of 400 µm and background doping concentration of $1 \times 10^{12}$ cm$^{-3}$) wafer—was placed on a temperature-controlled stage. The pump laser generated 200-fs pulses (Gaussian pulse duration and spatial profile) with a maximum energy of 500 µJ at a central wavelength of 515 nm, with a spot diameter about 3.6 mm (Fig. S3, *Supporting Information*). The repetition rate of the laser was 1 kHz, so the accumulation of heat across many pulses could be ignored (Sec. S5, *Supporting Information*). The s-polarized pump laser was incident at an angle of 45°. The generated thermal pulses were collected by a zinc-selenide lens in the normal direction, and focused onto a fast and broadband AC-coupled mid-infrared detector (Boston Electronics PVI-4TE-10.6, response time of 1 nanosecond with a bandwidth of 3 - 11 µm). Optional infrared filters could be inserted before the detector to obtain spectral selectivity. The output voltage from the detector was then recorded by an oscilloscope (Tektronix TDS7404B, 4 GHz, 20 GS/s).

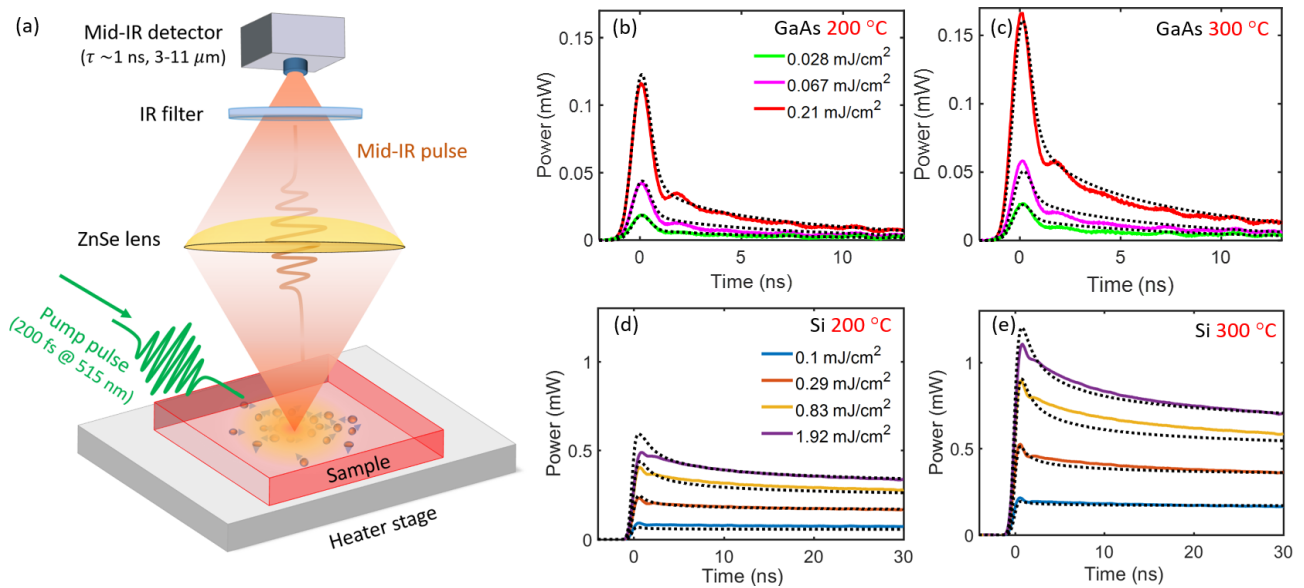

Figure 1: (a) Schematic of the experiment. A flat and undoped semiconductor wafer was placed on a temperature-controlled stage. 200-fs laser pulses at 515 nm with a maximum energy of 500 µJ were incident from 45° with a repetition rate of 1 KHz. Absorption of the pump-laser pulses generates and heats free carriers in the undoped semiconductor, resulting in thermal-emission pulses in the infrared. The thermal-emission pulses were collected with a zinc-selenide lens in the normal direction, and then detected using a fast mid-infrared detector (response time ~1 nanosecond, bandwidth of 3 to 11 µm). Optional infrared filters could be inserted

before the detector to obtain spectral selectivity. The solid lines in (b-e) show calibrated thermal-emission pulses from GaAs and Si with different pump fluences from an emitting area of 14 mm² towards a solid angle of 0.2 sr, at different stage temperatures. The data in (b-c) was collected without any filters. The corresponding theoretical calculations are shown with dotted black lines.

The measurements without spectral filtering for GaAs and Si with different heater-stage temperatures and pump fluences are shown in Fig. 1(b-e). Experimental data using higher pump-laser fluence, achieved by focusing the pump laser, are shown in Fig. S1 of the *Supporting Information*. The experiment using Si was performed with a higher pump fluence compared to the GaAs experiment because of the lower linear absorption coefficient of Si. Throughout this paper, we plot the estimated experimental power emitted toward the lens over the detector bandwidth of 3-11 μm; the conversions required to obtain these values are described in detail in *Supporting Information* Sec. 3.

Both thermal pulses detected from GaAs and Si show two features [Fig. 3(b-e)]: a one-nanosecond peak matching the temporal response of the detector is observed at the beginning of the pulse, and is followed by a much slower decrease of the falling edge of the pulse. For GaAs, a relatively stronger one-nanosecond peak and a faster decay of the falling edge were observed. Comparing the experimental data at different temperatures, the emission of both GaAs and Si increases by roughly a factor of two when the sample temperature is increased from 200 to 300 °C, consistent with the Stefan-Boltzmann law.

To better understand the experimental results and the process of thermal-pulse generation via free-carrier dynamics, we built a two-part model to calculate the time-dependent thermal emission from optically pumped semiconductors. The first part of the model characterizes the material response to an external optical pump pulse and the second part calculates the corresponding thermal emission.

Following the absorption of an optical pump pulse, the free-carrier density and temperature inside of a bulk semiconductor are functions of both depth and time. Therefore, we modeled the semiconductor wafer as a thin-film stack, with each layer having different time-dependent material properties and temperatures [Fig. 2(a)]. To simulate the material response under the optical pump, we adapted the model from Ref. [24], described in detail in Sec. 3 of the *Supporting Information*. This model characterizes the interaction between a semiconductor and an external optical field. The semiconductor is assumed to have certain linear and nonlinear absorption. The free-carrier optical and thermal properties, as well as lattice thermal properties, are also input parameters for this model. The interactions between light, carriers, and the lattice are simulated via coupled differential equations which we solve using the finite-difference time-domain method. The outputs of this portion of the model are the time- and depth-dependent free-carrier density and temperature and the lattice temperature.

Figure 2(b) shows the calculated free-carrier temperature as a function of depth for the top one micron of a GaAs wafer illuminated by a 200-fs pump pulse at 515 nm with a fluence of 0.21 mJ/cm², matching our experimental conditions. The pump pulse center was incident on the material surface at $t_0 = 0.5$ ps. The corresponding carrier density and lattice temperature are shown in Fig. S6 of the *Supporting Information*. The free carriers can be heated up to very high temperatures (~ 4000 K) for a very short time (< 3 picoseconds) due to the absorption of the visible-frequency pump. Heating of the free carriers occurs because the pump photon energy (2.41 eV) is much higher than the bandgap of GaAs (1.52 eV). This excess energy of the free carriers immediately after they are generated leads to a temperature that is much higher than that of the lattice. The peak temperature $T$ can be estimated using $T \sim (hf - E_g)/3k_B$, where $hf$ is the pump photon energy and $E_g$ is the bandgap. After the pump, the free carriers quickly reach thermal equilibrium with the lattice through electron-phonon interactions.

Long after the pump pulse (hundreds of picoseconds; Fig. S7 of the *Supporting Information*), the carriers are in complete equilibrium with the lattice. After this point, further changes of the free-carrier and lattice temperatures can be neglected because the lattice cooling time is on the order of microseconds (Sec. S5, *Supporting Information*), which is much longer than our measurement time. Therefore, the only significant change to the material system is the free-carrier density, which is determined by diffusion and recombination [24]:

$$\frac{\partial n}{\partial t} = -\gamma n^3 - \frac{n}{\tau_c} + D\Delta n. \tag{1}$$

Here $n, \gamma, \tau_c$ and $D$ represent the carrier density, Auger recombination coefficient, free-carrier lifetime, and ambipolar diffusion coefficient, respectively.

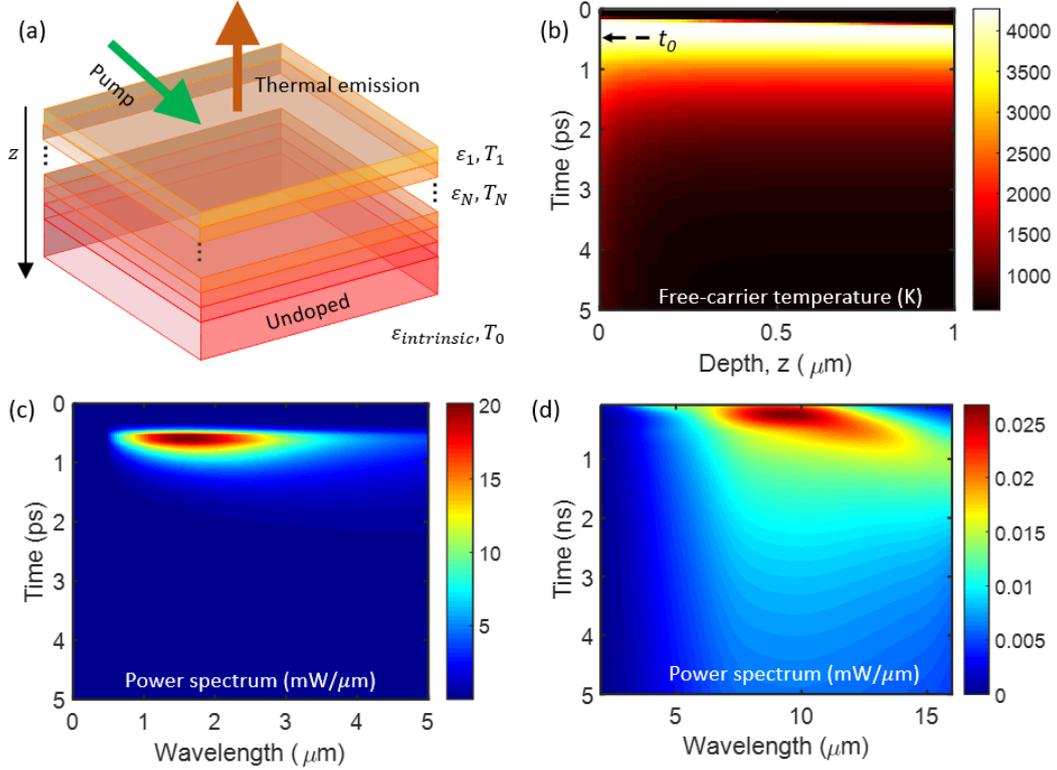

Figure 2. (a) The optically-pumped semiconductor wafer is modeled as a dynamic one-dimensional multilayer system, with each layer having different depth- and time-dependent material properties $\varepsilon(z,t)$ and temperature $T(z,t)$. (b) Calculated free-carrier temperature for a 300 °C GaAs wafer with pump fluence of 0.21 mJ/cm². The corresponding power spectrum of the calculated emitted (c) picosecond thermal pulse from hot carriers and (d) nanosecond thermal pulse from free-carrier-density variation. In (d), the time axis ranges from 100 ps to 5 ns after the pump pulse, such that no hot-carrier emission is seen. The calculation assumes a 14-mm² emitting area and a measurement solid angle of 0.2 sr, matching the experiments.

The impact of the free carriers on the optical properties can be described using the Drude model [22]:

$$\varepsilon(z,t) = \varepsilon_{intrinsic} - \omega_p(z,t)^2 / \left( \omega^2 + \frac{i\omega}{\tau} \right). \tag{2}$$

Here $\varepsilon$ and $\varepsilon_{intrinsic}$ are the permittivities of the photo-excited and intrinsic material, respectively, $\tau$ is the free-carrier relaxation time, and $\omega_p(z,t) = \sqrt{n(z,t)e^2/(m\varepsilon_0)}$ is the plasma frequency. $m$ is the free-carrier effective mass, which is related to the electron and hole masses $m_e$, $m_h$ via $m^{-1} = m_e^{-1} + m_h^{-1}$ [22]. Due to electron-electron and electron-lattice interactions, the free-carrier relaxation time depends on both the density [25], [26] and temperature [27], [28]. To account for this effect, we used the Caughey-Thomas relation that was originally formulated for the density-dependent mobility in silicon [29] and has also been used for GaAs [25] (Sec. S3, *Supporting Information*).

The second part of the model calculates the thermal emission from the time-dependent system. Kirchhoff's law cannot be used here directly because a single temperature cannot be defined: (1) the free-carrier and lattice temperatures can be different, and (2) both the free-carrier and lattice temperatures are depth-dependent (Fig. S5 and S6 of the *Supporting Information*). Therefore, to calculate thermal emission, we use a model based on the fluctuation-dissipation theorem (FDT) and dyadic Green's functions from Ref. [30] (Sec. 3, *Supporting Information*). Thermal emission originates from the thermally-induced random currents inside an object. The power spectrum of the random current sources is given by the FDT, and the Green's function can used to calculate the corresponding electromagnetic fields from these sources. In our one-dimensional thin film stack, the Green's function can be obtained via scattering matrix [30]. We note that the recently developed local Kirchhoff law [31] could also be used.

Figure 2(c, d) shows the calculated time-dependent power spectrum of the picosecond (due to the rapid

temperature change of the hot carriers) and nanosecond (due to carrier density dynamics, no hot-carrier contribution) features of the thermal emission from a GaAs wafer in the normal direction with a solid angle of 0.2 sr and an emitting area of 14 mm². The hot-carrier thermal pulse plotted in Fig. 2(c) is only a few picoseconds long due to the small hot-carrier lifetime. The spectral peak of this pulse is in the near-infrared region (around 1.5 μm), corresponding to the high temperature of the carriers [Fig. 2(b)]. The nanosecond thermal pulse plotted in Fig. 2(d) is much wider because the free-carrier lifetime is three orders of magnitude larger than the hot-carrier lifetime (nanoseconds vs. picoseconds). The spectral peak of the nanosecond thermal pulse is in the mid infrared (shifts from 8 μm to longer than 10 μm with increasing time), a result of the much-lower temperature of the free carriers when they are thermally equilibrated with the lattice (~600 K), as compared to immediately after the pump (~4000 K).

We performed calculations using our experimental parameters (Sec. 4, *Supporting Information*), and the results are plotted using dotted black lines in Figs. 1(b-e). The model parameters for Si and GaAs were chosen from reasonable values in the literature, with several of the parameters selected to achieve good agreement with the experiments (Sec. 3, *Supporting Information*). The calculations reproduce our experimental features quite well. With the help of the model, we are able to better understand our experimental results. The hot-carrier thermal emission results in an ultrafast pulse of only a few picoseconds wide [Figs. 2(c) and S9(a)], which leads to the one-nanosecond-duration peak experimentally observed in Fig. 1(b-e). The broader pulse in the experiment is due to the 1-ns response time of our detector. The much-slower decrease of the emission signal after the short hot-carrier pulse is due to the reduction of the photogenerated free-carrier density and thus the emissivity. GaAs has a free-carrier lifetime of a few nanoseconds, so the detected thermal-emission signal decreases to a very low value within 10 nanoseconds [Fig. 1(b, c), Fig. S1(a, b)]. The decay rate of the emission from Si is much lower than from GaAs [Fig. 1(d, e) and Fig. S1(c, d)]; Si is an indirect-bandgap semiconductor and has a much longer free-carrier lifetime (a few microseconds [21]). Within a temporal window of 30 nanoseconds, diffusion and Auger recombination are the primary mechanisms for changes to the carrier density for Si. Indeed, an increasing decay rate of the measured thermal-emission signal with increasing pump fluence is observed here, demonstrating the impact of Auger recombination.

Because thermal emission depends on temperature, the temperature of the stage-heated wafer should have a significant impact on the emitted pulses. In Figs. 1(b, c) and S1(a, b), the amplitude of the pulse resulting from the change in emissivity (the slower decay) increases by about a factor of 2, in agreement with a theoretical calculation that considers the Stefan-Boltzmann law. The one-nanosecond peak amplitude, however, does not change significantly with stage temperature, because this peak comes from hot-carrier thermal emission. As shown in our calculation [Fig. 2(b)], the free-carrier temperature is mainly determined by the pump photon energy and the bandgap rather than the wafer temperature.

As seen from Fig. 2(d), the spectrum of the thermal pulse changes with time due to the time-dependent emissivity and temperature. To determine the spectral composition of the thermal pulses, we performed measurements with four infrared transmission filters with different passbands placed between the sample and the detector [Fig. 1(a)]. The transmission windows for the filters are centered around 3, 4, 5, and 10 μm [Fig. 3(a)]. The thermal pulses from a 300 °C Si wafer (from an emitting area of 1.4 mm² into a solid angle of 0.2 sr) were measured through each filter with three different pump fluences (0.83, 2.14, and 6.18 mJ/cm²), and are plotted using solid lines in Fig. 3(b-d). The corresponding theoretical calculations are plotted using dotted black lines. Similar experimental data but using GaAs are shown in Fig. S2 in the *Supporting Information*.

When the pump fluence is low (Fig. 3(b), 0.83 mJ/cm²), the measurements with filters indicate that the emitted pulse has much more power at longer wavelengths. This is expected because the wafer temperature is only 300 °C, and corresponds to a peak thermal-emission wavelength in the mid-infrared region. When the pump fluence is increased (Fig. 3(c), 2.14 mJ/cm²), a relative increase in signal through the shorter-wavelength filters (3, 4 and 5-μm filters) is seen at the beginning of the pulse near $t = 0$ ns, indicating an increased contribution from emitting components at a higher temperature than the wafer and sample stage. This increase becomes obvious for the highest pump fluence (Fig. 3(d), 6.18 mJ/cm²): the emission measured through the 3-μm filter is the highest and is much higher than that through the 10-μm filter. This observation is evidence of thermal emission from hot carriers. The hot carriers quickly reach thermal equilibrium with the lattice (Fig. S7, *Supporting Information*); as seen in Fig. 3(d), the emission measured through shorter-wavelength filters decays very quickly after the pump pulse. Long after the pump, the strongest emission is through the 10-μm filter.

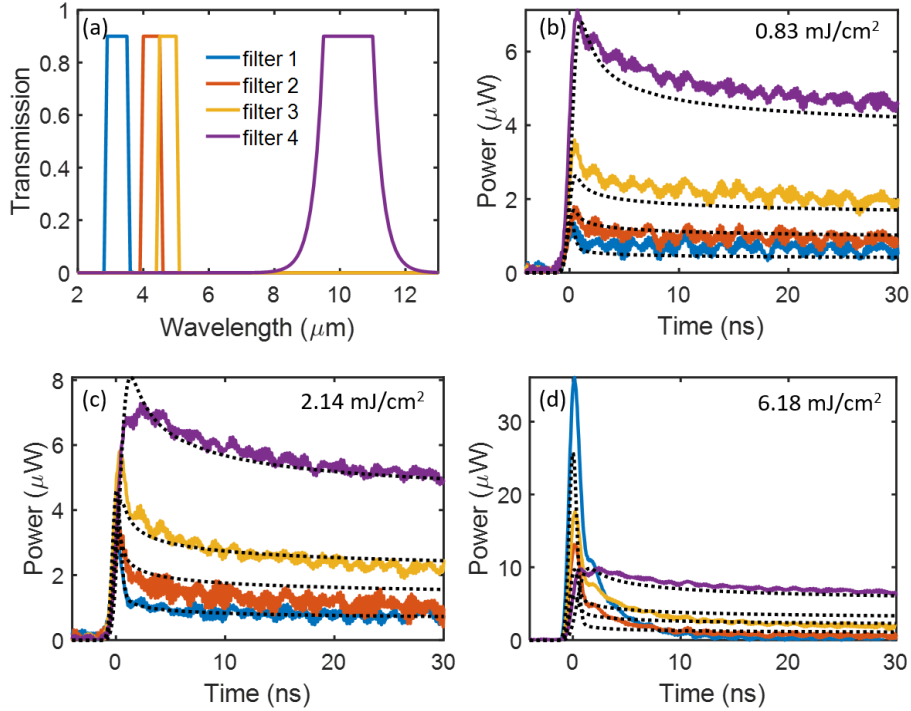

Figure 3. (a) Transmission spectra of the infrared filters used in our experiment. (b, c, d) Solid lines: experimentally calibrated detected thermally emitted power from a 1.4 mm² emitting area into a solid angle of 0.2 sr, using the filters, from a Si wafer at 300 °C with pump fluence of (b) 0.83, (c) 2.14, and (d) 6.18 mJ/cm². The corresponding theoretical calculations are shown using dotted black lines.

To conclude, we experimentally demonstrated the generation of few-nanosecond thermal pulses via emissivity modulation of semiconductor surfaces using an ultrafast visible pump. We also observed sub-nanosecond thermal emission from hot carriers in semiconductors. One can view this process of generating mid-infrared pulses from ones in the visible as an unconventional type of frequency conversion, which differs substantially from traditional nonlinear processes. For example, the conversion efficiency from the visible to the mid infrared not only depends on the pump power, but also on the sample temperature. Though the resulting pulse energies are limited, this method of generating mid-infrared pulses has substantial tunability compared to existing approaches. The temporal and spectral features of this thermal pulse, which depend on free-carrier dynamics, can be tuned by choosing different combinations of the pump laser and semiconductor material, and the bandwidth and directionality can be engineered using various nanophotonic design structures [5]; for example using gratings [32] or by patterning the pump-laser beam profile [15].

The authors would like to thank Dr. Lynda E. Busse from the Naval Research Laboratory for kindly providing us the filters, Josh Ostrander from Dr. Martin Zanni's group for helping test our infrared detector, and Jonathan L. King for discussions on heat-transfer simulations. Y.X., J.S. and M.K. are supported by the Office of Naval Research (ONR) under Grant No. N00014-16-1-2556. N.C. and A.P. are supported by the ONR through the Naval Research Laboratory Basic Research Program.

Supporting Information:

# Nanosecond mid-infrared pulse generation via ultrafast-modulated thermal emissivity

Yuzhe Xiao[1], Nicholas A. Charipar[2], Jad Salman[1], Alberto Piqué[2] and Mikhail A. Kats[1,3,4,*]

[1]Department of Electrical and Computer Engineering, University of Wisconsin-Madison, Madison, Wisconsin, USA 53706
[2] Naval Research Laboratory, 4555 Overlook Ave. SW, Washington, DC 20375, USA
[3]Materials Science and Engineering, University of Wisconsin-Madison, Madison, Wisconsin, USA 5370
[4]Department of Physics, University of Wisconsin - Madison, Madison, Wisconsin 53706, USA
*Email address: mkats@wisc.edu

## S1: Additional experimental data

We measured pulsed thermal emission from silicon (Si) and gallium arsenide (GaAs) wafers with a higher pump fluence than that shown in the manuscript by focusing the pump laser spot size down to 1.4 mm$^2$, which is 10% of the original spot area. This increases the fluence by a factor of 10 for the GaAs case. For Si, we found that 10× the fluence from Fig. S1 resulted in damage, so we reduced the laser power somewhat; the resulting fluences are in the legend of Fig. S1. In Fig. S1, we plot the experimentally calibrated emitted power, radiated into a solid angle of 0.2 sr within the detector bandwidth of 3-11 μm, for stage temperatures of 200 and 300 °C. The corresponding calculations from our theoretical model are plotted using dotted black lines.

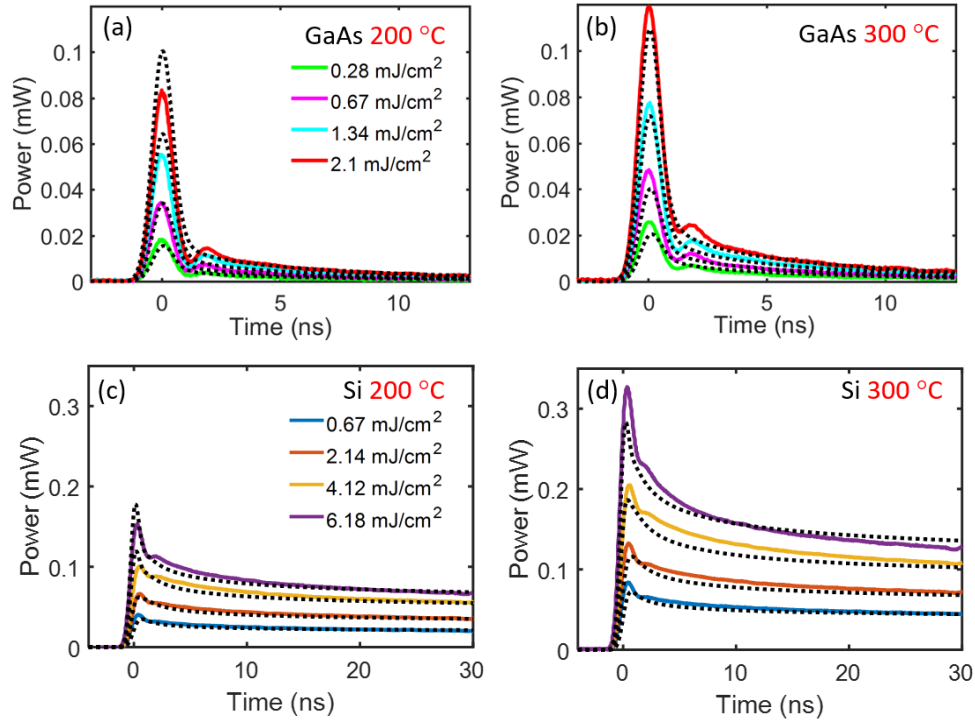

Figure S1. Solid lines: experimentally calibrated thermal-emission power from polished (a, b) GaAs and (c, d) Si from an emitting area of 1.4 mm$^2$ into a solid angle of 0.2 sr, at stage temperatures of 200 and 300 °C, for higher pump fluences than those shown in the main text. The corresponding theoretical calculations are shown with black dotted lines.

We also used filters to obtain spectral information about the thermal pulses from the GaAs wafer, just like for Si in the main text (Fig. 3). Figure S2 shows the calibrated experimental emitted power passing through each filter (Fig. 3(a) in the main text) from a 300 °C GaAs wafer with an emitting area of 1.4 mm$^2$ into a solid angle of 0.2 sr, for pump fluence of 0.67, 1.34, and 2.1 mJ/cm$^2$. The corresponding calculations from theoretical model are also shown.

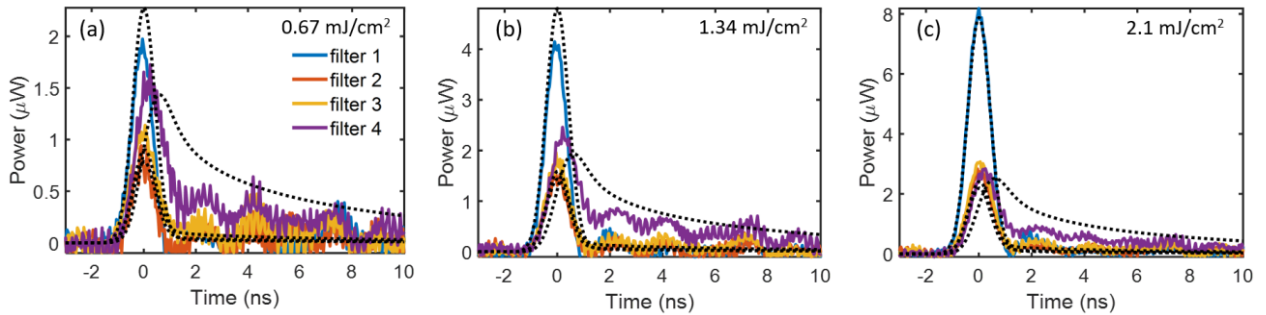

Figure S2. Solid lines: Measured emitted power (from a 1.4 mm² emitting area into a solid angle of 0.2 sr) through different filters (shown in Fig. 3(a), with passbands near 3, 4, 5 and 10 μm for filter 1, 2, 3 and 4, respectively) from a 300 °C GaAs wafer with pump fluence of (a) 0.67, (b) 1.34, and (c) 2.1 mJ/cm². The corresponding theoretical calculations are shown using dotted black lines.

## S2: Additional details of the experimental setup

Here, we present additional details of the experiment setup, including the pump laser profile and an estimation of the detection efficiency (the fraction of power that is detected divided by the total power passing through our collection lens).

The intensity profile of the 515-nm pump laser pulse is shown in Fig. S3, which can be approximated by a Gaussian profile with an effective diameter (width of a beam that is 4 times σ, where σ is the standard deviation of the intensity distribution) of about 3.6 mm. When the beam is incident at an angle of 45°, the corresponding illuminated (and hence emitting) area on the sample is about 14 mm².

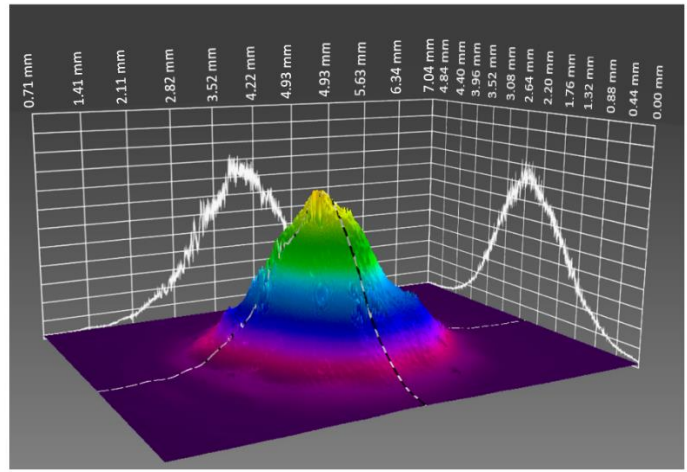

Figure S3. Measured intensity profile of the 515-nm pump pulse laser at normal incidence (Spiricon SP620U). The laser spot can be approximated by a Gaussian with an effective diameter of about 3.6 mm.

We used a single zinc selenide (ZnSe) lens (diameter of 25.4 mm, focal length of 25 mm) to collect and focus the thermal emission onto the detector. The lens was placed about 60 mm away from the sample in the normal direction, corresponding to a collection solid angle of about 0.2 sr. The two sample emitting areas were 14 and 1.4 mm² for the unfocused and focused case, respectively. The active area of the detector is 0.25 mm² (Boston Electronics PVI-4TE-10.6, 0.5 mm × 0.5 mm).

Since only a fraction of the emitted power that passes through the lens is incident on the active area of the detector, the precise emitted power was not directly measurable. To estimate the actual power emitted by the sample into a certain solid angle, we performed ray-tracing simulations using Zemax OpticsStudio to approximate the total power passing through the lens (Fig. S4). Figure S4(a) shows the layout of the case with the unfocused pump. For the unfocused case, the emitting source area is 14 mm². In this case, based on simulations, only about 3.5% of emitted power was incident on the detector element [Fig. S4 (b)]. In the focused case, the emitting area is smaller, and more-easily images onto the active area of the detector, resulting in 15% of the emitted light being focused onto the detector element [Fig. S4 (c)]. This increase in collection efficiency matches our experimental observations.

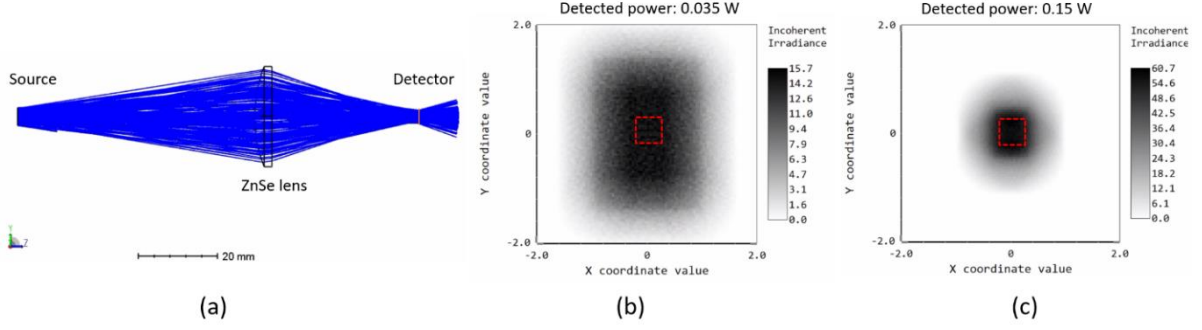

(a)                       (b)                       (c)

Figure S4. (a) Layout of the ray-tracing simulation using Zemax OpticsStudio for the unfocused case: the source has an emitting area of 14 mm² and was placed 60 mm away from the lens. The source is about 1.4 times larger in in the y-direction than x-direction, to mimic the oval shape of the pump laser spot on the sample. The detector has an active area of 0.25 mm² and was placed about 40 mm after the lens. (b) Detector analysis for the unfocused case: about 0.035 W of power was incident on the detector element (rectangular area) assuming 1 W emitting from the source towards the lens (corresponding to a solid angle of about 0.2 sr). (c) Detector analysis for the focused case: the source area is reduced from 14 to 1.4 mm². About 0.15 W of power was incident on the detector element (rectangular area), assuming 1 W emitting from the source towards the lens.

Because the actual power emitted in a given solid angle is more fundamental than the power detected by our finite-size detector element, the pulse power we report in all experimental figures (Figs. 1, 3, S1 and S2) is the estimated power emitted toward the lens in the detector bandwidth of 3 to 11 µm, which is the actual power detected divided by the corresponding fraction ratio estimated from ray tracing simulations. The detected power is obtained from the detected voltage through the detector voltage responsivity ($1.4 \times 10^4$ V/W).

## S3: Theoretical modeling

This section shows more details of the model we used to calculate the temporal and spectral features of pulsed thermal emission, including the details of light-matter interaction inside our two semiconductors, the scattering rates of the free carriers used in the Drude model, and the method of calculating thermal emission from an object with spatially and temporally varying temperature and optical properties.

We adopted the model from Ref. [S1] to calculate the electron-hole pair (free-carrier) density and temperature, as well as the lattice temperature when a semiconductor is excited by an optical pulse. More specifically, the following equations are solved:

$$\frac{\partial n}{\partial t} = \frac{\alpha I(z,t)}{h\nu} + \frac{\beta I^2(z,t)}{2h\nu} - \gamma n^3 - \frac{n}{\tau_c} + \theta n - \nabla \cdot \boldsymbol{J} \tag{S1}$$

$$\boldsymbol{J} = -D\left(\nabla n + \frac{n}{2k_B T_e}\nabla E_g + \frac{n}{2T_e}\nabla T_e\right) \tag{S2}$$

$$\boldsymbol{W} = \left(E_g + 4k_B T_e\right)\boldsymbol{J} - (k_e + k_h)\nabla T_e \tag{S3}$$

$$C_{e-h}\frac{\partial T_e}{\partial t} = (\alpha + \theta n)I(z,t) + \beta I^2(z,t) - \frac{C_{e-h}}{\tau_{hot}}(T_e - T_l)$$

$$-\nabla \cdot \boldsymbol{W} - \frac{\partial n}{\partial t}\left(E_g + 3k_B T_e\right) - n\left(\frac{\partial E_g}{\partial T_l}\frac{\partial T_l}{\partial t} + \frac{\partial E_g}{\partial n}\frac{\partial n}{\partial t}\right) \tag{S4}$$

$$C_l\frac{\partial T_l}{\partial t} = \nabla \cdot (k_l\nabla T_l) + \frac{C_{e-h}}{\tau_{hot}}(T_e - T_l) \tag{S5}$$

$$\frac{\partial I}{\partial z} = -(\alpha + \theta n)I(z,t) - \beta I^2(z,t) \tag{S6}$$

The free-carrier density $n$ is solved using Eq. S1, where $I$ is the light intensity, $\alpha, \beta$ are linear and two-photon absorption coefficients, $h$ is Planck's constant, $\nu$ is pump light frequency, $\gamma$ is the Auger recombination coefficient, $\tau_c$ is the free-carrier lifetime, and $\theta$ is the impact ionization coefficient. The free-carrier current $J$ and ambipolar energy $W$ are defined using Eqs. S2-3, where $T_e, T_l$ are the free-carrier and lattice temperatures, $E_g$ is the bandgap, $k_e, k_h$ are the thermal conductivity of electrons and holes. The free-carrier and lattice temperature are solved using Eqs. S4-5, where $C_{e-h}, C_l$ are the heat capacitances for electron-hole pairs and the lattice, and $\tau_{hot}$ is the lifetime of the hot carriers. Changes to light intensity $I$ is described by Eq. S6, where $\Theta$ is the free-carrier absorption cross-section. Note that $n, I, J, W, T_e$, and $T_l$ are all depth- ($z$) and time-dependent, where $z$ is the distance into the material from the surface. Solving these equations using a one-dimensional finite-difference time-domain (FDTD) algorithm can yield the depth- and time-dependent free-carrier density, free-carrier temperature, and the lattice temperature.

Table S1 summarizes all the parameters that we used when solving Eqs. S1-6 for Si and GaAs. Most of the model parameters were chosen according to the reported literature values. Some of the parameters ($\tau_{hot}, \tau_c, \gamma$) are chosen to have the best agreement with experimental results, and are close to the reported literature values. Since the model was built for Si [S1], not all of the parameters are readily available for GaAs, which is why we had to make some assumptions below. This also might make our model less reliable for GaAs.

Table S1: Model parameters

| Properties | Silicon | Gallium Arsenide |
|---|---|---|
| $k_l$ [W/(cm·K)] | $1585 T_l^{-1.23}$ [S1] | $686 T_l^{-1.25}$ [S2] |
| $C_l$ (J/cm³) | $1.978 + 3.54 \times 10^{-4} T_l - T_l^{-2}$ [S1] | $1.72 + 2.8 \times 10^{-4} T_l - 1.87 \times 10^3 T_l^{-2}$ [S3] |
| $k_e$ [eV/(s·A·K)] | $-3.47 \times 10^8 + 4.45 \times 10^6 T_e$ [S1] | $-2.1 \times 10^9 + 2.7 \times 10^7 T_e$ [S1][*] |
| $\tau_{hot}$ (fs) | $500 \cdot (1 + n/6 \times 10^{20}\ \mathrm{cm}^{-3})$ [S4]–[S6] | $750 \cdot (1 + n/6 \times 10^{20}\ \mathrm{cm}^{-3})$ [S7][S8] |
| $\gamma$ (cm⁶/s) | $10 \times 10^{-31}$ [S1][S9] | $11 \times 10^{-30}$ [S10] |
| $\tau_c$ (ns) | 1000, Appendix G of Ref. [S11] | 10, Appendix G of Ref. [S11] |
| $\theta$ ($s^{-1}$) | $3.6 \times 10^{10} \exp(-1.5 E_g/k_B T_e)$ [S1] | (assumed to be negligible) [S12] |
| $D_0$ (cm²/s) | $18 \cdot (T_{rm}/T_l)$ [S1] | $20 \cdot (T_{rm}/T_l)$ [S13] |
| $E_g$ (eV) | $1.16 - 7.02 \times \frac{10^{-4} T_l^2}{T_l + 1108} - 1.5 \times 10^{-8} n^{1/3}$ [S1] | $1.52 - 8.87 \times \frac{10^{-4} T_l^2}{T_l + 572} - 2 \times 10^{-11} n^{-1/2}$ [S14], [S15] |
| $\alpha$ (cm⁻¹) | $3.42 \times 10^4$ [S16] | $7.81 \times 10^4$ [S17] |
| $\beta$ (cm/GW) | (assumed to be negligible) [S1] | 3.6 [S18][S19][+] |
| $\Theta$ (cm²) | $5.1 \times 10^{-18} \cdot (T_l/T_{rm})$ [S1] | $5.5 \times 10^{-18} \cdot (T_l/T_{rm})$ [S20][o] |
| $m$ (free electron mass) | 0.15 [S11] | 0.058 [S11] |

[*] Value derived from Si, assuming the thermal conductivity of electrons is proportional to its mobility.

[+] Using the experimental value near 1 μm from ref. [S19] , and the scaling rule from ref. [S18] to extrapolate the value for 0.515 μm.

[o] Using the experimental value near 1 μm.

The following expression was used for the scattering time of free carriers in the Drude model [S21]:

$$\tau = \frac{\tau_{max} - \tau_{min}}{1 + (n/N_0)^\alpha} + \tau_{min}. \tag{S7}$$

Note that immediately after the pump, the carriers have energy much higher than the bandgap (hot carriers), which increases their interaction with the lattice, thereby increasing their scattering rate [S22], [S23]. Due to the additional scattering from the lattice, values for the time constants for the hot carriers are an order of magnitude smaller than those for carriers that have cooled down by reaching thermal equilibrium with the lattice [S22], [S23]. The parameters used in Eq. S7 in our model are chosen according to literature values [S24], [S25] and are shown in Table S2.

Table S2: Model parameters for the free-carrier scattering time

| Properties | Silicon | | Gallium Arsenide | |
|---|---|---|---|---|
| | Hot carriers | Cold carriers | Hot carriers | Cold carriers |
| $\tau_{max}$ (fs) | 4 | 200 | 7.5 | 375 |
| $\tau_{min}$ (fs) | 0.4 | 20 | 0.1 | 5 |
| $N_0$ (1/cm³) | $0.5 \times 10^{17}$ | $0.5 \times 10^{17}$ | $1 \times 10^{17}$ | $1 \times 10^{17}$ |
| $\alpha$ | 0.7 | 0.7 | 0.2 | 0.2 |

To calculate the thermal emission from a system with depth- and time-dependent temperature and permittivity distributions, we used a model from Ref. [S26]. Consider the thermal radiation emitted by an arbitrary object with position-dependent temperature $T(\boldsymbol{r'})$ and permittivity $\epsilon(\boldsymbol{r'})$ evaluated at an observation point $\boldsymbol{r}$. The position-dependent temperature results in a position-dependent distribution of random currents $\boldsymbol{j}(\boldsymbol{r'}, t)$ and their Fourier transform, $\boldsymbol{j}(\boldsymbol{r'}, \omega)$ throughout the object. The power spectrum of these random currents is given by the fluctuation-dissipation theorem (FDT). The simplified FDT, valid for isotropic linear materials described by dielectric constant $\epsilon(\omega)$, is (Chapter 14 of Ref. [S27]):

$$\frac{1}{\omega Im[\epsilon_0\epsilon(\omega)]}\pi < j_\alpha(\boldsymbol{r_1},\omega)j_\beta^*(\boldsymbol{r_2},\omega') >= \Theta(\omega,T)\delta(\boldsymbol{r_1}-\boldsymbol{r_2})\delta(\omega-\omega')\delta_{\alpha\beta}, \tag{S8}$$

where $\alpha$ and $\beta$ refer to different Cartesian components of the current, $\Theta(\omega, T) = \hbar\omega/(e^{\frac{\hbar\omega}{k_B T}} - 1)$ is the mean (expected) energy in a particular state with frequency $\omega$ given temperature $T$, $\epsilon_0$ is the vacuum permittivity, $\delta$ is the Dirac delta function, and $\delta_{\alpha\beta}$ is the Kronecker delta. The deltas express that the fluctuating currents at different positions, frequencies, and orthogonal directions are uncorrelated. The ensemble average ($<...>$) is the average over all combinations of possible random currents.

The averaged radiative intensity at $\boldsymbol{r}$, $< \boldsymbol{S}(\boldsymbol{r}, \omega) >$, where $\boldsymbol{S}$ is the Poynting vector, is related to the averaged spectral power of the random current source $< j_\alpha(\boldsymbol{r}, \omega)j_\beta^*(\boldsymbol{r}, \omega) >$, which is given by FDT. Therefore, in order to determine the radiated intensity, one just needs to find the relation between the source current and the corresponding fields. For a current density at some point in space, $\boldsymbol{j}(\boldsymbol{r'}, t)$, the resulting fields at $\boldsymbol{r}$ is given by the dyadic Green's function of the system $\bar{G}(\boldsymbol{r}, \boldsymbol{r'})$ (Chapter 14.7 of Ref. [S28]):

$$\boldsymbol{h}(\boldsymbol{r}, \boldsymbol{\omega}) = \iiint \boldsymbol{j}(\boldsymbol{r'}, \omega) \cdot \bar{G}(\boldsymbol{r}, \boldsymbol{r'}, \omega) \, dv', \tag{S9}$$

where $\boldsymbol{h}$ is the electric or magnetic field for the corresponding electric and magnetic Green's function. The averaged radiative intensity $< \boldsymbol{S}(\boldsymbol{r}, \omega) >$ is determined both by the Green's function $\bar{G}(\boldsymbol{r}, \boldsymbol{r'})$ and spectral power of the random currents $< j_\alpha(\boldsymbol{r}, \omega)j_\beta^*(\boldsymbol{r}, \omega) >$. In our one-dimensional thin film stack, the Green's function can be obtained via the scattering matrix [S26]. The calculation based on FDT and Green's function is performed for different times so that we finally have time-dependent thermal emission.

## S4: Results from the model

Figure S5 shows the calculated free-carrier density ($n$), carrier temperature ($T_e$) and lattice temperature ($T_l$) for a Si wafer on a temperature stage at 300 °C, illuminated by a 200-fs Gaussian pump pulse with two different fluences: 0.1 and 1.92 mJ/cm². The pump pulse center arrives at the material surface at $t_0 = 0.5$ ps. As shown in Fig. S5 (a), a peak carrier density about $2 \times 10^{19}$ cm⁻³ is generated near the surface with a low pump fluence of 0.1 mJ/cm². The carrier density shows an exponential decay into the material, with most of the carriers well confined to an approximately 1-µm-thick layer. After the pump (t > 1 ps), the carrier density remains almost unchanged because of the long free-carrier lifetime. When the pump fluence is increased to 1.92 mJ/cm² [S5 (d)], the peak carrier density increases to $4 \times 10^{20}$ cm⁻³. However, unlike in the lower-intensity pump scenario, the carrier density, particularly near the surface ($z = 0$ µm), decreases quickly after the pump, due to Auger recombination.

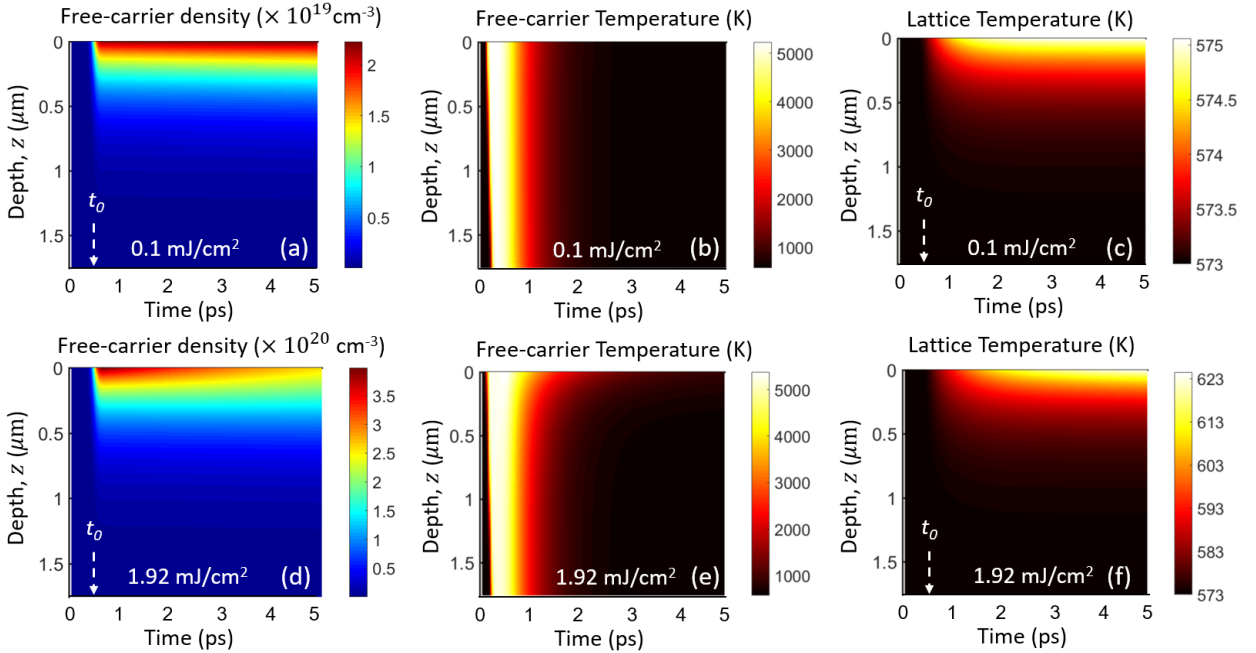

Figure S5. Calculated time- and depth-dependent free-carrier density (a and d), free-carrier temperature (b and e), and lattice temperature (c and f) for a 300 °C (537 K) Si wafer excited by a 200-fs Gaussian pump at 515 nm with fluence of 0.1 (top) and 1.92 mJ/cm² (bottom).

As shown in Fig. S5(b, e), the free carriers are heated to a very high temperature, far out of equilibrium with the lattice immediately after the pump. The peak free-carrier temperatures are at about 5000 K for both pump fluences [Fig. S5 (b, e)], which agrees with the estimated peak temperature of 4830 K using $T \sim (hf - E_g)/3k_B$, where $hf - E_g$ is the difference between the pump photon energy and the band gap, and $3k_B T$ is the corresponding thermal energy of the carrier. Due to the effect of Auger heating [S29], the peak free-carrier temperature is higher and the free-carrier cooling time is longer for the higher pump case (e), especially near the material surface ($z \sim 0$ µm). For both (c) low and (f) high pump fluences, the lattice temperature increases due to the interaction with the hot carriers. Heating of the lattice is negligible for a pump fluence of 0.1 mJ/cm², while the lattice temperature can be increased by more than 50 K when the pump fluence is raised to 1.92 mJ/cm². The lattice and electrons reach full thermal equilibrium within a few picoseconds in low-fluence case, while the duration is much longer (100 ps) for the strong-pump case [Fig. S7 (a)]. The lattice temperature gradually decreases from the surface into the material. The impact of such a temperature gradient on the actual thermal emitted power is discussed in Sec. S6.

Figure S6 shows the calculated $n$, $T_e$, and $T_l$ for GaAs with a pump fluence of 0.028 (top) and 0.21 mJ/cm² (bottom). Two major differences between GaAs and Si are worth noting. First, the linear absorption coefficient is larger for GaAs, leading to a smaller penetration depth. Therefore, the generated free carriers are confined in a narrower region near the surface for GaAs [Fig. S6(a, d)]. Second, GaAs has a larger bandgap than Si. Therefore, the peak hot-carrier temperature of GaAs [~4000 K, Fig. S6(b, e)] is lower than that for Si.

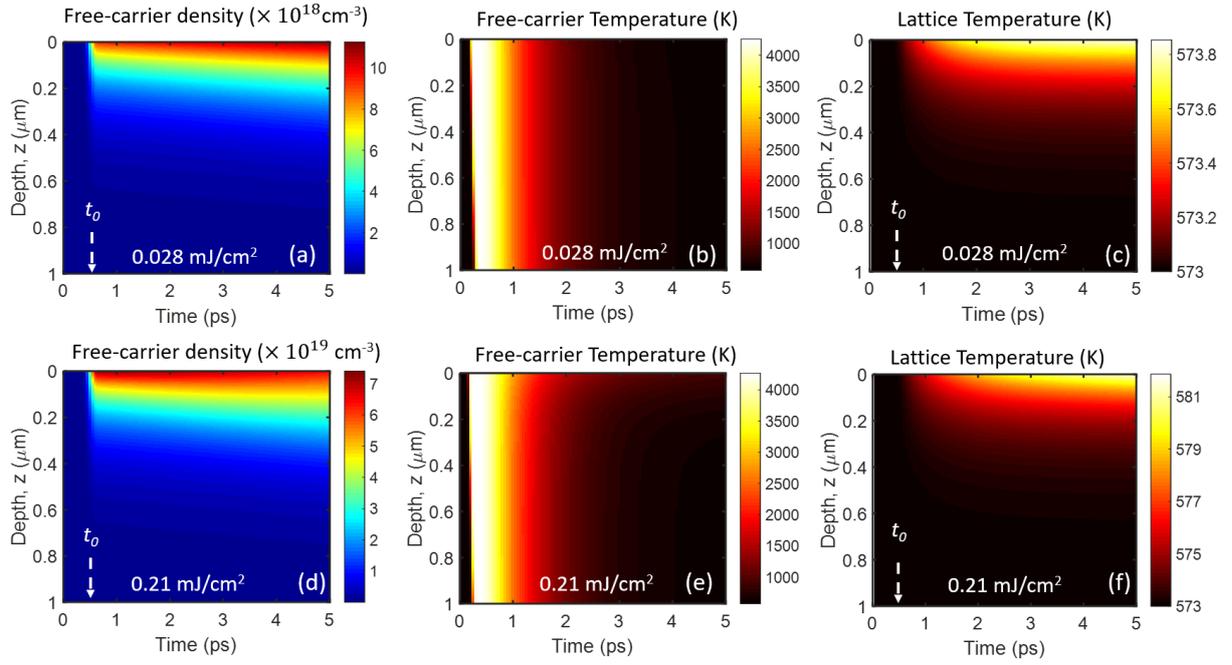

Figure S6. Calculated time- and depth-dependent free-carrier density (a and d), free-carrier temperature (b and e), and lattice temperature (c and f) for a 300 °C GaAs wafer excited by a 200-fs Gaussian pump at 515 nm with fluence of 0.028 (top) and 0.21 mJ/cm² (bottom).

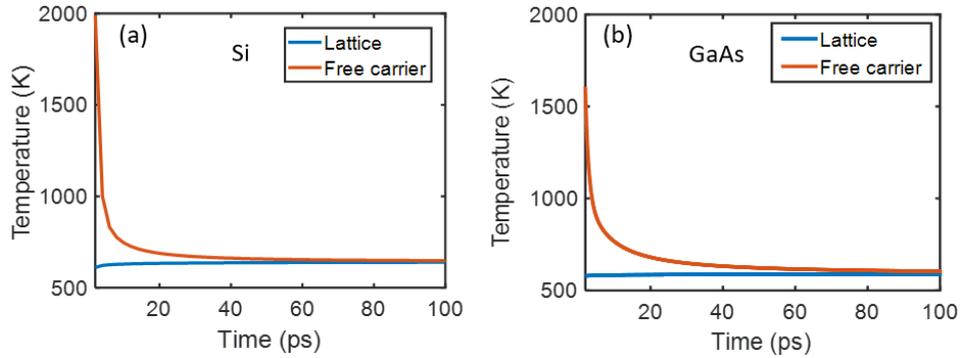

Figure S7. (a) Calculated lattice (blue) and free-carrier (red) temperature at the surface ($z = 0$ μm) for a 300 °C Si wafer excited by a 200-fs Gaussian pump at a central wavelength of 515 nm with fluence of 1.92 mJ/cm². (b) Same as (a), but with a GaAs wafer with pump fluence of 0.21 mJ/cm².

Figure S8(a) shows the calculated free-carrier density for Si [Fig. S5 (d)] after the pump pulse, from 0.25 to 30 ns. Shortly after the pump, the free carriers are confined within a < 1-μm-thick layer. The highly confined free carriers quickly diffuse within a few nanoseconds, causing the carrier density to significantly decrease. After 20 ns, the change of carrier density over time becomes much smaller. This is partially due to the relative weaker effect of diffusion, but mainly due to the slow free-carrier recombination of Si (carrier lifetime of a few μs). Figure S8(b) shows the case for GaAs [Fig. S6 (d)]. The carrier dynamics for GaAs within the first 5 ns are similar to that of Si. However, due to the fast carrier recombination for GaAs (lifetime of a few nanoseconds), the carrier density keeps decreasing even when the effect of carrier diffusion becomes very small (at $t > 10$ ns).

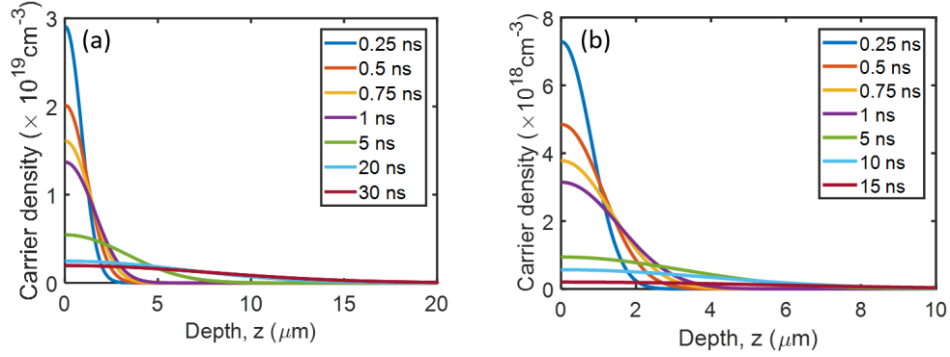

Figure S8. Calculated free-carrier density distribution at different times after the pump for (a) a Si wafer with pump fluence of 1.92 mJ/cm², and (b) a GaAs wafer with pump fluence of 0.21 mJ/cm².

Once the free-carrier temperature and the material properties are known, the corresponding thermal emission can be calculated. Here we show the integrated thermal-emission power within the spectral range from 3 to 11 μm (matching our detector wavelength range) for a 300 °C GaAs wafer with pump fluence of 0.21 mJ/cm². The emitting area is 14 mm² and the solid angle over which we integrate is 0.2 sr. Figure S9 shows the calculated thermal emission from hot carriers, after the carriers have thermalized (after 100 ps), and the total power that should be detected by our experimental setup. The total detected power in Fig. S9(c) is obtained by convolving (a) and (b) with the temporal response function of our detector (Boston Electronics PVI-4TE-10.6, 0.5 mm × 0.5 mm), which is a Gaussian function with a width of 1 ns.

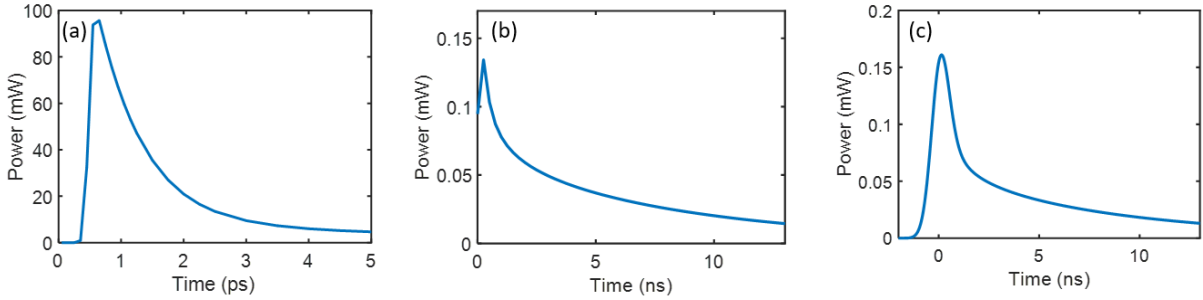

Figure S9. Calculated integrated pulsed thermal emission within the spectral range from 3 to 11 μm (matching our detector bandwidth) for a 300 °C GaAs wafer with pump fluence of 0.21 mJ/cm². (a) and (b) show the emission from hot carriers and after the carriers have thermalized (the plot starts at time = 100 ps), respectively. The emitting area is 14 mm² and we integrate over a solid angle of 0.2 sr. The calculated total signal from the detector is shown in (c).

### S5: Estimation of the conductive cooling rate of the lattice

As shown in the simulation [Figs. S5(c) and S6(c)], the heating of the lattice due to optical pumping is negligible when the pump fluence is low. However, there can be non-negligible laser-induced heating of the lattice when the pump power is sufficiently high. Indeed, our model predicts increases of the lattice temperature by 50 K for Si with pump fluence of 1.92 mJ/cm² [Fig. S5 (f)], and by 9 K for GaAs with pump fluence of 0.21 mJ/cm² [Fig. S6 (f)].

To verify that the nanosecond modulation of thermal emission measured comes from the emissivity change due to free-carrier dynamics and not from temperature changes of the lattice, we estimated the cooling rate of the heated wafer by performing heat-transfer simulations. Figure S10 shows the schematic of the structure with some initial temperature distribution due to the pump. We model a thin layer of material representing the laser-heated portion ($\Delta z \sim 0.5$ for GaAs and 1 μm for Si, according to the calculated results from Fig. S5 and S6) on top of a very thick layer of material representing the remaining wafer.

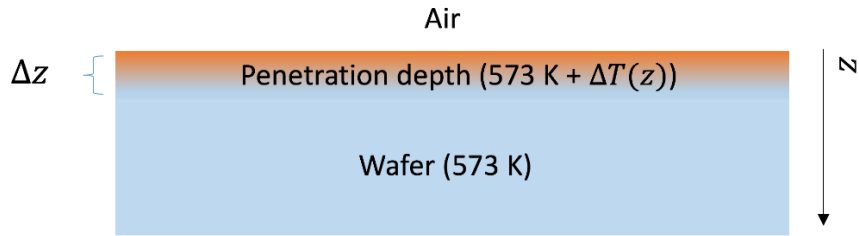

Figure S10. Initial temperature distribution after pump-laser illumination, for heat-transfer simulations.

The lattice temperature of the structure was calculated using the one-dimensional heat equation:

$$\frac{\partial T(t,z)}{\partial t} = \frac{k}{\rho C_p} \frac{\partial^2 T(t,z)}{\partial z^2} \tag{S10}$$

where $k, \rho,$ and $C_p$ represent the thermal conductivity, density, and heat capacity of the material, respectively. For the structure shown in Fig. S10, two cooling channels exist: (1) conductive cooling into the substrate and (2) convective cooling into the surrounding air. The thermal conductivities and heat capacity for Si and GaAs we used are in Table S1, while the density are shown in Table S3 (taken from Appendix G of Ref. [11]). The initial temperature distribution was chosen according to the lattice temperature at the end of the FDTD simulation at 100 ps for Si and GaAs.

Table S3: Heat-transfer parameters

| Properties | Silicon | Gallium Arsenide |
|---|---|---|
| $\rho$ (kg/m³) | 2330 | 5317 |

Equation S10 was solved for a $L = 10\ \mu m$ thick wafer, with a convective heat-transfer boundary condition at $z = 0\ \mu m$ and a fixed temperature $T = 573\ K$ at $z = 10\ \mu m$, which corresponds to the temperature of the stage. Figure S11 (a, b) shows the calculated temperature distribution of Si and GaAs for the first 30 and 15 ns, respectively. As shown here, the changes in the lattice temperature over this time scale can be neglected.

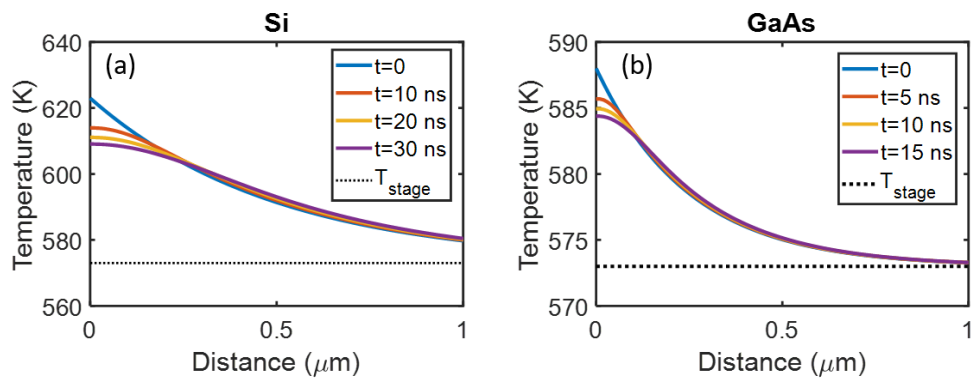

Figure S11. Calculated temperature of the wafer as a function of the distance from the surface for (a) Si with a pump fluence of 1.92 mJ/cm² and (b) GaAs with a pump fluence of 0.21 mJ/cm².

Further evidence that lattice temperature change does not play a significant role comes from the comparison in the experiment data between Si and GaAs in Figs. 1 and S1. GaAs and Si have similar thermal properties, and therefore similar cooling dynamics, but the decay rate of the thermal-emission signal in these experiments is very different between these two materials. Therefore, the nanosecond-scale modulation of thermal emission is clearly a result of the change in emissivity, rather than the temperature.

Figure S12 shows the calculated temperature distributions in the Si and GaAs wafers after a significantly longer period of time. As shown here, the entire substrate cools down after about 100 µs for Si and 200 µs for GaAs. The laser repetition rate is 1 kHz in our experiment, corresponding to a separation time between adjacent pulses of 1 ms. Therefore, the accumulated heating effect due to repetitive pulses can be neglected.

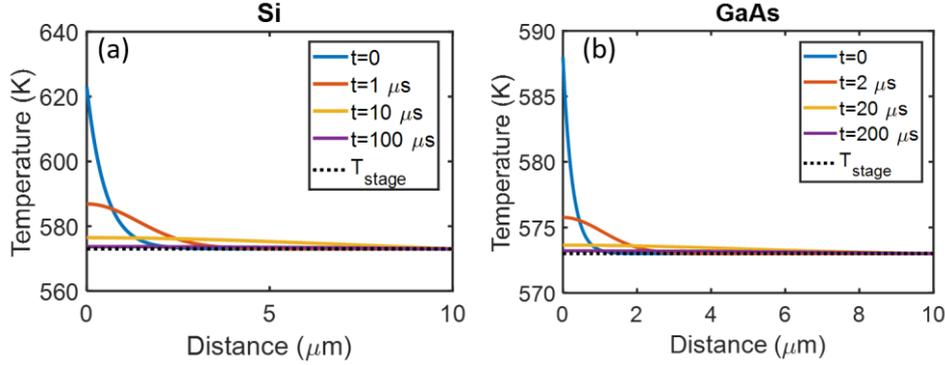

Figure S12. Calculated lattice temperature as a function of wafer depth after a longer period has elapsed post pumping for (a) Si and (b) GaAs.

### S6: Thermal emission from objects with depth-dependent temperature distributions

The intensity of the pump laser is attenuated as it propagates into the material. Due to the gradient pump intensity, the heating of the free carriers and the lattice is not uniform [Figs. S5(f) and S6(f)]. For a system with a non-uniform temperature distribution, the thermally emitted power cannot be trivially calculated via Kirchhoff's Law, $i.e.$, by multiplying the blackbody radiation distribution at a particular temperature by the emissivity. The situation is even more complicated immediately after the pump, when the free carriers at a much higher temperature than the lattice. Instead, we calculate the thermal emission directly, using the method described in Section S3, to obtain the emitted power from such a system. Note that here the optical properties are also depth-dependent.

Figure S13 compares the theoretical thermally emitted power from a GaAs wafer long after the pump is extinguished ($t > 0.1$ ns, so no hot carriers), calculated both directly and using Kirchhoff's law, for pump fluences of 0.028, 0.21 and 2.1 mJ/cm² to mimic our experimental conditions. In these calculations, an emitting solid angle of 0.2 sr and an emitting area of 14 mm² is assumed. In the Kirchhoff's-law calculation, the temperature of the surface ($z = 0$ µm) at $t_0 = 100$ ps (when free carriers are in thermal equilibrium with the lattice) was used, while in the direct calculation, the actual lattice temperature at each depth was used. The lattice temperatures for the pump fluence 0.028 and 0.21 mJ/cm² are shown in Fig. S6(c, f), respectively, while the lattice temperature for the highest pump fluence of 2.1 mJ/cm² is shown in Fig. S13 (d).

As shown in Fig. S13(a), the indirect and direct ways of calculating thermally emitted power are identical for the lowest pump fluence. This is expected since the gradient in the lattice temperature is minimal with a total temperature difference of less than 1 K, or less than 0.2 % of the relative temperature change [Fig. S6(c)]. The effect of the gradient temperature distribution on the emitted power becomes noticeable as the pump fluence is increased to 0.21 mJ/cm², and the temperature difference between the surface and substrate is ~ 10 K. As a result, the emitted power from the direct calculation is about 5% smaller than that from Kirchhoff's law. This effect becomes very significant for the highest pump fluence of 2.1 mJ/cm², where the temperature drop is more than 100 K. In this case, the directly calculated emitted power is about 50% of the value calculated indirectly via Kirchhoff's law.

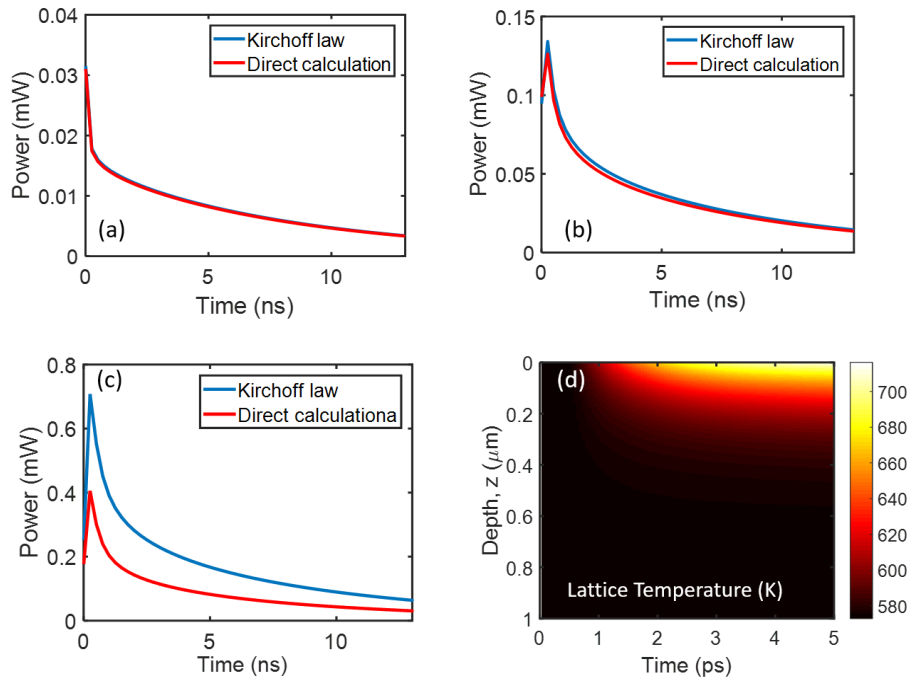

Figure S13. Calculated thermally emitted power from a GaAs wafer long after the pump ($t > 0.25$ ns), using Kirchhoff's law (blue curve, using the surface lattice temperature) and direct calculation (red) with pump fluence of 0.028, 0.21 and 2.1 mJ/cm², respectively. (d) The lattice temperature for a 300 °C GaAs wafer excited by a 200 fs Gaussian pump at 515 nm with a fluence of 2.1 mJ/cm².